\documentclass[12pt,preprint]{aastex}
\usepackage{color}
\shorttitle{Prompt GeV-TeV Emission from Proton-Dominated GRBs}
\shortauthors{Asano, Inoue \& M\'esz\'aros}

\input{colordvi.tex}

\begin{document}

\title{
Prompt High-Energy Emission from Proton-Dominated Gamma-Ray Bursts}
\author{\scshape Katsuaki Asano\altaffilmark{1},
Susumu Inoue\altaffilmark{2}, and
Peter M\'esz\'aros\altaffilmark{3}}
\email{asano@phys.titech.ac.jp, inoue@tap.scphys.kyoto-u.ac.jp, nnp@astro.psu.edu}

\altaffiltext{1}{Interactive Research Center of Science, 
Tokyo Institute of Technology, 2-12-1 Ookayama, Meguro-ku, Tokyo 152-8550, Japan}
\altaffiltext{2}{Department of Physics, 
Kyoto University, Oiwake-cho, Kitashirakawa, Sakyo-ku, Kyoto 606-8502, Japan}
\altaffiltext{3}{Department of Astronomy \& Astrophysics;
Department of Physics;
Center for Particle Astrophysics;
Pennsylvania State University,
University Park, PA 16802}

\date{Submitted; accepted}

\begin{abstract}
The prompt emission of gamma-ray bursts (GRBs)
is widely thought to be radiation from accelerated electrons,
but an appreciably larger amount of energy could be carried by accelerated protons,
particularly if GRBs are the sources of ultra-high-energy cosmic rays (UHECRs).
We model the expected photon spectra for such ``proton-dominated'' GRBs in the internal 
shock scenario through Monte Carlo simulations,
accounting for various processes related to high-energy electrons and protons.
Besides proton and muon synchrotron components, emission from photomeson-induced 
secondary pair cascades becomes crucial, generally enhancing the
GeV-TeV and/or eV-keV photons and offering a signature of UHE protons.
In some cases, it can overwhelm the primary electron component
and result in GRBs peaking in the 10 MeV - 1 GeV range, which may be relevant 
to some bursts discussed in a recent re-analysis of EGRET TASC data.
The dependence of the spectra on key quantities such as the bulk Lorentz factor, 
magnetic field and proton-to-electron ratio is nontrivial due to the nonlinear 
nature of cascading and the interplay of electron- and proton-induced components.
Observations by {\it Fermi}, ground-based telescopes and other facilities
should test these expectations and provide critical constraints on the proton 
acceleration efficiency.
\end{abstract}

\keywords{cosmic rays --- gamma rays: bursts --- gamma rays: theory --- radiation mechanisms: nonthermal}

\maketitle

\section{Introduction}
\label{sec:intro}

The prompt emission of gamma-ray bursts (GRBs) is believed to arise
from ultrarelativistic outflows with bulk Lorentz factors $\Gamma \ga100$
\citep[see, e.g., reviews by][]{pir05,mes06}.
In the popular internal shock model,
collisions among inhomogeneities within the flow lead to formation of shocks
that convert a fraction of the bulk kinetic energy into Fermi-accelerated relativistic electrons,
whose synchrotron emission powers the observed MeV-band gamma-rays \citep{ree94}.
Initially, most of the kinetic energy as well as the internal energy generated via shock dissipation
are likely carried by protons,
so such models entail the operation of a physical mechanism
that transfers energy from protons to electrons on sufficiently short timescales.
This presumably occurs via collective electromagnetic processes,
as simple Coulomb collisions may be too slow.
A general problem in collisionless shock theory and GRB models in particular
is that this mechanism is poorly understood,
and one must frequently resort to a phenomenological parametrization.
In view of the large observed energy in MeV gamma rays,
the efficiency of proton-to-electron energy transfer is usually considered to be high.
However, this is by no means physically guaranteed.
In the case of supernova remnant shocks,
the total energy in accelerated electrons is often constrained observationally
to be much less than in protons \citep[e.g.][]{aha06}.
Since we do not yet understand the nature and total energy budget of the central engine,
we cannot readily exclude the possibility
that GRBs actually contain a significantly larger amount of energy in protons
compared to that radiated by the accelerated electrons.

Furthermore, a natural expectation is that the shocked protons are also Fermi-accelerated.
The physical conditions in internal shocks
may allow maximum energies $\ga 10^{20}$ eV,
so GRBs are potential sources of the observed ultra-high-energy cosmic rays
\citep[UHECRs;][]{wax95,vie95,mil96}.
The total energy in accelerated protons that must be supplied per burst
depends on a number of uncertain factors (see also App. B of \citet{mur08}).
The required local UHECR emissivity at proton energy $\varepsilon_p \sim 10^{19}$ eV
is $\varepsilon_p^2 d\dot{N}_p/d\varepsilon_p \simeq 0.8 \times 10^{44}\ {\rm erg\ Mpc^{-3} yr^{-1}}$ \citep{wax98,der07}.
Post-{\it SWIFT} estimates of the local rate of long GRBs
range from $0.2-1\ {\rm Gpc^{-3} yr^{-1}}$ if the GRB rate is proportional to the star formation rate,
down to $\sim 0.05\ {\rm Gpc^{-3} yr^{-1}}$ if the GRB rate evolves more strongly with redshift,
which may be observationally favored \citep[e.g.][]{dai06,le07,gue07}.
Assuming a power-law proton spectrum with index $p_p=2$,
the necessary isotropic-equivalent energy per burst
in accelerated protons integrated over $\varepsilon_p \sim 10^9 - 10^{20}$ eV
is $E_{\rm p} \sim 2 \times 10^{54} - 3 \times 10^{55}$ erg,
which is approximately independent of the actual beaming factor. 
Steeper spectra and hence even larger $E_{\rm p}$ are called for
if GRBs also contribute significantly to CRs below $10^{19}$ eV \citep{wic04}.
To be compared is the corresponding energy in accelerated electrons $E_{\rm e}$,
which can be roughly equated with the observed, isotropic-equivalent
MeV gamma-ray energy $E_{\gamma,{\rm iso}}$,
typically $\sim 10^{53}$ erg and up to $\sim 10^{54}$ erg
in the $1-10^4$ keV rest-frame band \citep{koc08}.
Thus, in order for GRBs to be viable sources of UHECRs,
the latest observations point to a highly proton-dominated energy budget,
$E_{\rm p}/E_{\rm e} \ga 10 - 100$.
The observed heterogeneity of GRBs also suggests that
not all bursts may be equally efficient UHECR accelerators,
in which case even higher $E_{\rm p}/E_{\rm e}$ may be warranted
for a subset of the bursts.

It is therefore of great interest whether such ``proton-dominated'' GRBs
can be diagnosed observationally.
A promising window is GeV-TeV gamma-rays,
where distinctive signatures of UHE proton acceleration may show up,
such as synchrotron emission from protons, muons or
secondary particles injected via photomeson interactions
\citep[e.g.][hereafter AI07]{vie97,boe98,gup07,asa07}.

AI07 recently undertook a detailed investigation of such emission processes
utilizing a comprehensive Monte Carlo code.
However, having assumed that the accelerated protons do not carry excessive extra energy,
their study was restricted to $E_{\rm p}/E_{\rm e}=1$.
In view of the above possibilities, here we follow and extend the work of AI07 to $E_{\rm p}/E_{\rm e} > 1$.
The results, which are often qualitatively and drastically different from AI07,
are discussed in relation to existing and upcoming observations.
Note that high-energy emission from proton-dominated GRBs
has been discussed previously in different contexts \citep[e.g.][]{tot98,asa03}.

After a recap of our formulation in \S \ref{sec:model},
we discuss the results and their observational implications
in \S \ref{sec:results} and \S \ref{sec:obs}, respectively,
and conclude in \S \ref{sec:conc}.

\section{Model and Methods}
\label{sec:model}

We briefly summarize the model and methods of AI07,
which should be consulted for more details.
In accord with the internal shock paradigm,
the emitting region corresponding to an individual pulse in the prompt light curve
is taken to be a homogeneous shell expanding with $\Gamma$ 
at radii $R$ from the central engine.
The comoving width of the shell is $l=R/\Gamma$  
and the pulse timescale in the observer frame is $\Delta t=R/\Gamma^2 c$,
as long as $R$ exceeds the shell spreading radius \citep{mes93},
which is always the case here.
Shock dynamics and time variability are not explicitly treated,
so our results should be interpreted as the time-averaged spectra for each pulse.

With given injection of accelerated electrons and protons 
in magnetic field $B$, we solve self-consistently
for the distribution of particles and photons in the shell using Monte Carlo techniques.
The time steps are always taken to be sufficiently shorter
than the particle cooling timescales \citep{asa05}.
In addition to synchrotron and inverse Compton (IC) emission from all particles,
our code includes synchrotron self-absorption, cascade processes with
photon-photon ($\gamma \gamma$) production of electron-positron pairs ($e^\pm$)
and Klein-Nishina regime Compton scattering,
as well as proton-induced processes such as photomeson ($p\gamma$) interactions
and secondary pion, muon and pair injection.
We adopt experimental results for the cross sections of
$p \gamma \to n \pi^+$, $p \pi^0$, $n \pi^+ \pi^0$ and $p \pi^+ \pi^-$,
while $p \gamma \to p \pi^0 \pi^0$ is neglected in view of its small cross section.
In case the primary proton is converted to a neutron,
we assume that it continues to interact with photons in the shell
during the comoving expansion timescale $t_{\rm exp}=l/c$.
We do not acccount for the minor contribution from neutron-decay electrons \citep{raz06}.
More details on the treatment of meson production and their decay products
can be found in \citet{asa05} and \citet{asa06}.

Furthermore, we now account for
the Bethe-Heitler (BH) pair production process ($p \gamma \to p e^+ e^-$),
whose cross section and inelasticity are taken from \citet{cho92}.
In the present context, the proton energy loss is always dominated by photopion production,
and the huge compactness of GRBs implies that the resultant electromagnetic cascade emission
is not very sensitive to the details of particle injection at high energies.
Thus, compared to cases neglecting the BH process,
we find that its inclusion here only leads to modest enhancements
of the secondary photon emission, by at most a few tens of percent.

Primary electrons with total energy density $U_{\rm e}$ are injected
with a power-law distribution $n_{\rm e}(\gamma_{\rm e}) \propto \gamma_{\rm e}^{-p_{\rm e}}$
in the range of Lorentz factors $\gamma_{\rm e,min} \le \gamma_{\rm e} \le \gamma_{\rm e,max}$.
The balance of Fermi acceleration and radiative cooling timescales gives
$\gamma_{\rm e,max}$.
Likewise, protons with total energy density $U_{\rm p}$ are injected with a distribution
$n_{\rm p}(\gamma_{\rm p}) \propto \gamma_{\rm p}^{-p_{\rm p}}$
in the range $\gamma_{\rm p,min} \le \gamma_{\rm p} \le \gamma_{\rm p,max}$.
We obtain $\gamma_{\rm p, max}$
by equating $t_{\rm acc}= \gamma_{\rm p} m_p c^2/e B c$,
the Fermi acceleration time in relativistic shocks,
to $\min(t_{\rm exp}, t_{\rm loss})$, where
$t_{\rm loss}$ is the energy loss time due to synchrotron, IC and $p\gamma$ cooling \citep{asa05}.
In mildly relativistic internal shocks,
$\gamma_{\rm p, min}$ should be of order unity;
here we take $\gamma_{\rm p, min}=10$.

The injection index for electrons is fiducially chosen to be $p_{\rm e}=2.5$,
implying $\beta \simeq 2.25$ for the spectral index above the synchrotron peak energy.
This is consistent with the mean of the $\beta$ values measured by BATSE,
albeit with a considerable dispersion,
from $\beta \la 1.5$ to $\beta \ga 3.0$ \citep{pre00,kan06}.
For protons, our fiducial index is $p_{\rm p}=2.0$,
appropriate when GRBs contribute to UHECRs only above $10^{19}$ eV \citep{wax98};
steeper spectra would increase still the energy demands.
Note that the values of $p_{\rm e}$ and $p_{\rm p}$ relevant to our results
each correspond to very different energy ranges;
GeV-TeV for electrons and 10-100 PeV for protons in the comoving frame.
Although the injection spectra for the two species
are expected to be the same at low energies where their gyroradii overlap,
$p_{\rm e} > p_{\rm p}$ may be effectively realized
if the proton spectrum covering 7-8 decades in energy deviates from a pure power-law
and becomes concave.
This may plausibly occur due to
1) nontrivial geometry and wavelength distribution of magnetic turbulence at the shock \citep{nie06},
2) nonlinear back-reaction of CR pressure on the shock structure \citep{bar91,mal01},
or 3) superposition of pre-existing and newly-injected particles
originating from different regions in the outflow \citep{bos08}.
Nevertheless,  
in view of the observed spread in $\beta$
and the uncertainties associated with obtaining time-integrated spectra,
we also discuss cases with $p_{\rm e}=p_{\rm p}$ in \S \ref{sec:equal}.

Some combinations of the remaining parameters are constrained so as to reproduce
typically observed properties of the MeV emission.
For given $B$ and $\Gamma$,
$\gamma_{\rm e,min}$ is chosen
such that the observed synchrotron peak energy for nearby bursts is
$\varepsilon_{\rm pk} = \Gamma \gamma_{\rm e,min}^2 \hbar e B/m_{\rm e} c \simeq 300$ keV.
Since the fast-cooling, primary electrons radiate away most of their energy 
as MeV photons within $\Delta t$, 
$E_{\rm e} = (4\pi \Gamma^2 R^2 c \Delta t) U_{\rm e}\simeq (4 \pi R^3) U_{\rm e}$
can be identified with $E_{\rm sh}$,
the observable, isotropic-equivalent MeV pulse energy.

Instead of $U_{\rm e}$, $U_{\rm p}$ and $U_{\rm B}=B^2/8\pi$,
hereafter we use $\epsilon_{\rm e}$, $\epsilon_{\rm p}$ and $\epsilon_{\rm B}$,
the conventional parametrization of the corresponding energies
as fractions of the shock-dissipated internal energy \citep[e.g.][]{mes06}.
Thus $\epsilon_{\rm B}/\epsilon_{\rm e}=U_{\rm B}/U_{\rm e}$ and 
$\epsilon_{\rm p}/\epsilon_{\rm e}=U_{\rm p}/U_{\rm e}=E_{\rm p}/E_{\rm e}$.
In place of $R$, we choose the observable $\Delta t$ as a parameter
and set $\Delta t = 0.1$ s for simplicity.
Below we only show the spectra corresponding to single pulses.
For bursts composed of $N$ similar pulses,
the duration-integrated energy would be simply $N$ times larger,
$E_{\gamma,{\rm iso}}=N E_{\rm sh}$.
The set of parameters are then
$\Delta t$, $E_{\rm sh}$, $\Gamma$,
$\epsilon_{\rm B}/\epsilon_{\rm e}$ and  $\epsilon_{\rm p}/\epsilon_{\rm e}$.
All spectra are plotted as observed fluence versus photon energy,
assuming a GRB redshift $z=0.1$.
Spectral attenuation by intergalactic $\gamma\gamma$ absorption is neglected.

\section{Results}
\label{sec:results}

\subsection{Fiducial Spectral Indices}
\label{sec:fiducial}

\begin{figure}[htb!]
\centering
\epsscale{1.0}
\plotone{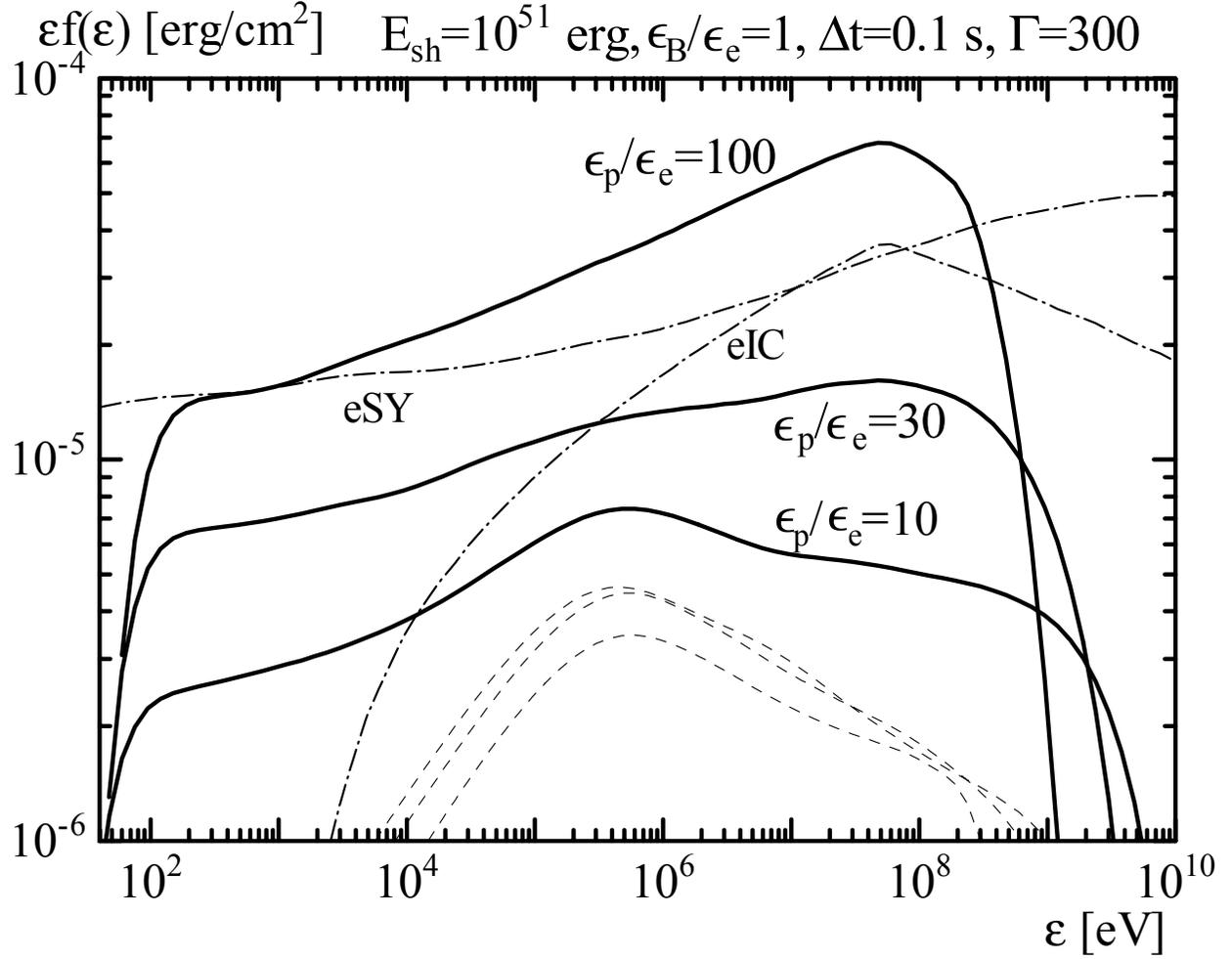}
\caption{
Single pulse, prompt photon spectra for varying 
$\epsilon_{\rm p}/\epsilon_{\rm e}$ as labeled.
Other parameters are marked above the figure.
Dashed curves denote the primary contribution only,
whose peak flux decreases with $\epsilon_{\rm p}$.
Dot-dashed curves denote separately the electron synchrotron (labeled eSY)
and inverse Compton (eIC) components
without $\gamma\gamma$-absorption effects
for $\epsilon_{\rm p}/\epsilon_{\rm e}=100$.
\label{fig:fp-dep}}
\end{figure}

First we discuss different cases with our fiducial values of $p_{\rm e}=2.5$ and $p_{\rm p}=2.0$.
As mentioned above, $\varepsilon_{\rm pk}$ is chosen to have the typically observed value of 300 keV.
Prompt emission spectra of single pulses
for $E_{\rm sh}=10^{51}$ erg, $\Gamma=300$, $\epsilon_{\rm B}/\epsilon_{\rm e}=1$
and varying $\epsilon_{\rm p}/\epsilon_{\rm e} =10-100$
are shown in Figure \ref{fig:fp-dep}.
The sharp spectral cutoffs at low and high energies
are due to synchrotron self-absorption
and $\gamma \gamma$ absorption, respectively.
This applies to all spectra below when such sharp cutoffs are seen.
Most remarkable is the prominent $e^\pm$ cascade component,
i.e. synchrotron and IC emission from secondary $e^\pm$
triggered by $p\gamma$ interactions of UHE protons with low energy photons.
For the lower range of $\epsilon_{\rm p}/\epsilon_{\rm e}$,
primary synchrotron photons constitute the main $p\gamma$ target.
However, as the proton content increases,
the target photons become dominated by
synchrotron emission from the low energy part of the secondary $e^\pm$ themselves.
The dependence of the spectra on $\epsilon_{\rm p}/\epsilon_{\rm e}$
is therefore nonlinear and not simply proportional, as apparent in Figure \ref{fig:fp-dep}.
The secondary photons also affect the primary synchrotron component
(dashed curves in Figure \ref{fig:fp-dep})
through enhanced IC cooling,
even though the injection distribution is unchanged.

In general, cascade emission significantly hardens the high-energy spectra.
Since secondary $e^\pm$ with Lorentz factors $< \gamma_{\rm e,min}$ can be 
injected in the cascade, it can also give rise to excess UV-to-X-ray 
emission lying above the extrapolation of the sub-MeV spectra, as seen 
for $\epsilon_{\rm p}/\epsilon_{\rm e}=10-30$ in Figure \ref{fig:fp-dep}.
The entire spectra thus tends to become flat in $\varepsilon f(\varepsilon)$.

The case of $\epsilon_{\rm p}/\epsilon_{\rm e}=100$ is drastically different.
Here the proton-induced secondary emission totally overwhelms any primary 
electron component, resulting in a hard spectrum peaking at 10-100 MeV.
Although approximately a single power-law between 100 eV and 30 MeV,
in fact it comprises two emission processes by secondary $e^\pm$,
mainly synchrotron $\la$ MeV and IC $\ga$ MeV (dot-dashed curves in Figure \ref{fig:fp-dep}).
Despite $\epsilon_{\rm B}/\epsilon_{\rm e}=1$,
IC can dominate over synchrotron since the energy density of secondary $e^\pm$
exceeds both $U_{\rm B}$ and $U_{\rm e}$.

\begin{figure}[htb!]
\centering
\epsscale{1.0}
\plotone{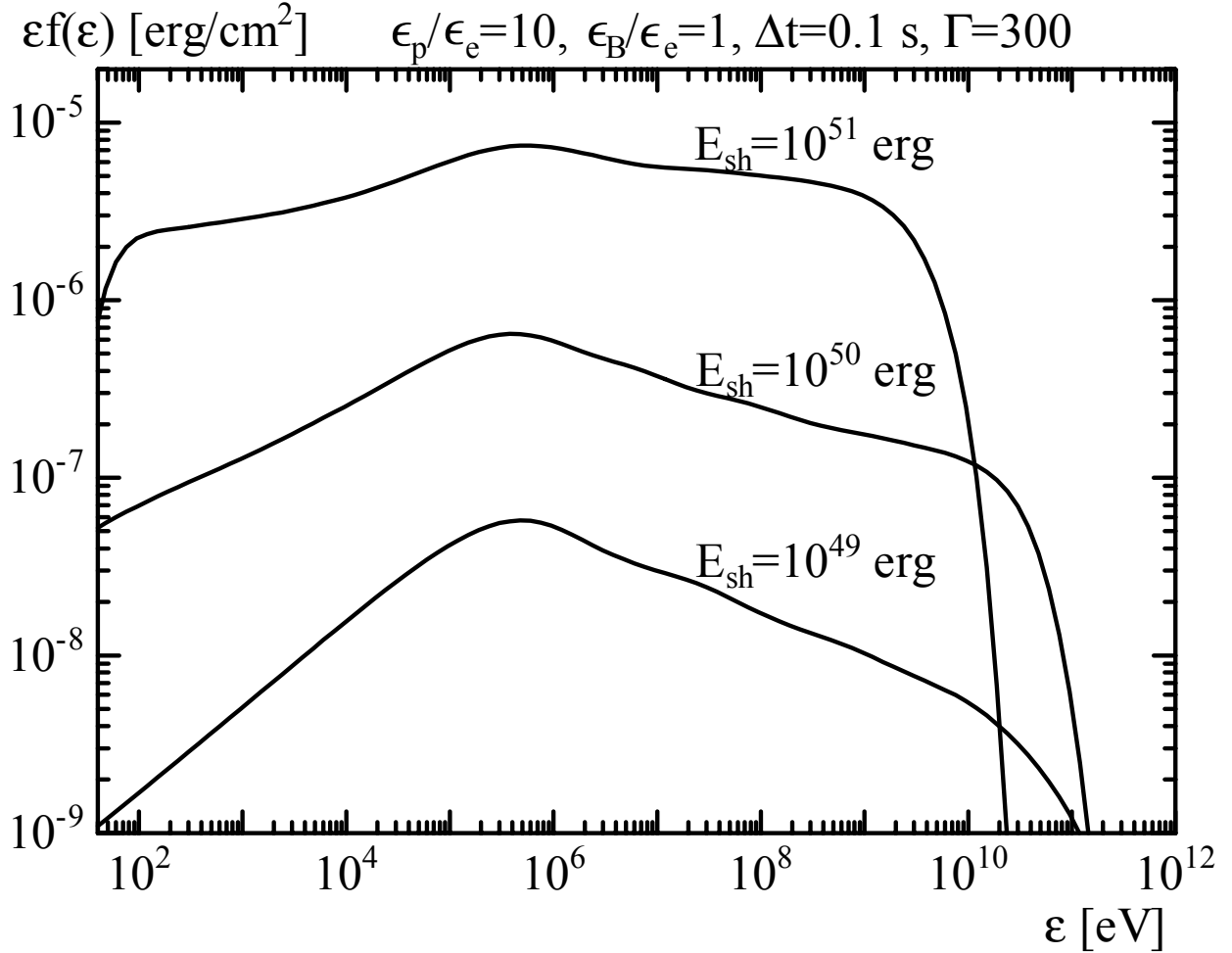}
\caption{
Single pulse, prompt photon spectra for
varying $E_{\rm sh}$ as labeled.
Other parameters are marked above the figure.
\label{fig:Esh-dep}}
\end{figure}

\begin{figure}[htb!]
\centering
\epsscale{1.0}
\plotone{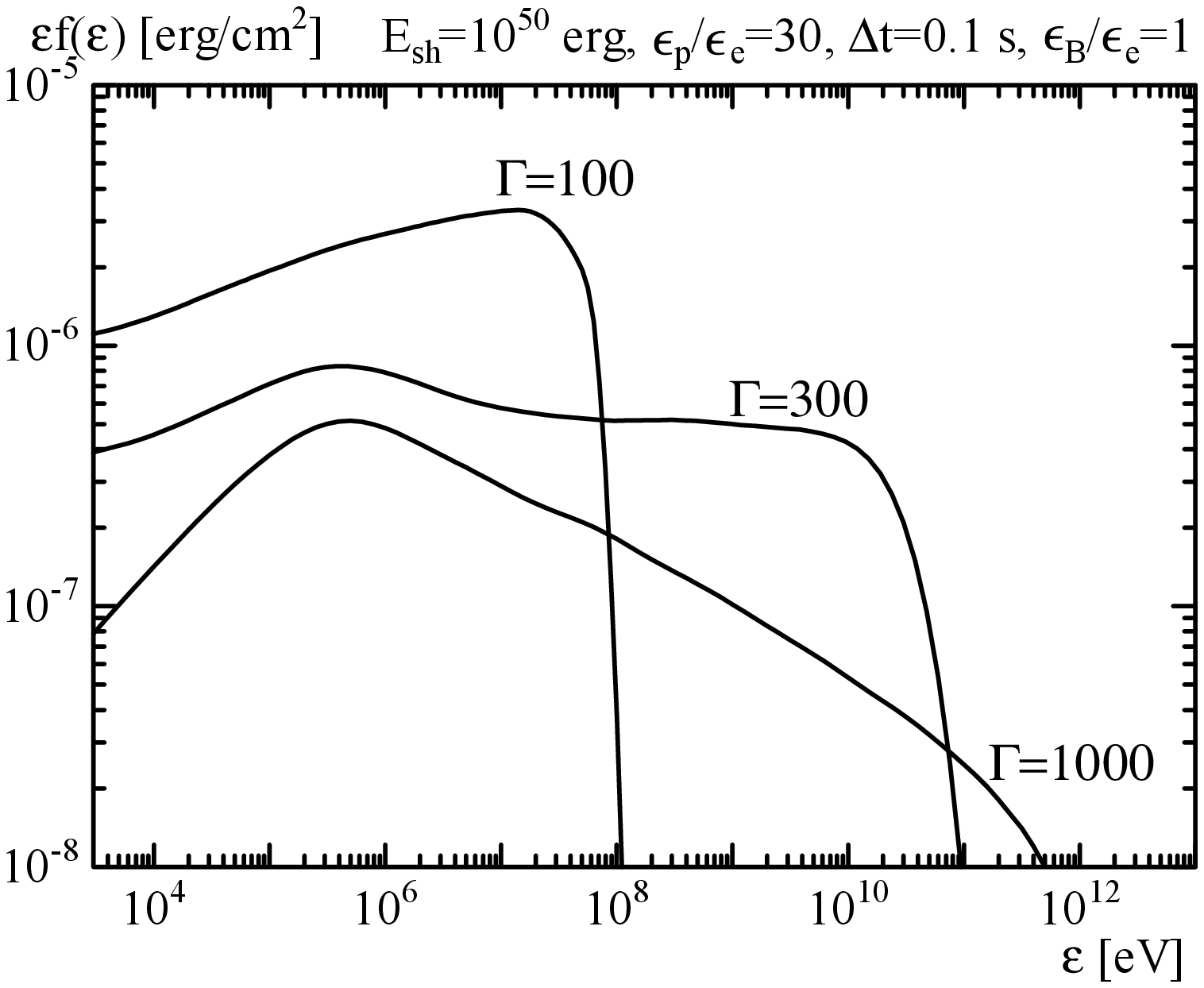}
\caption{
Single pulse, prompt photon spectra for
varying $\Gamma$ as labeled.
Other parameters are marked above the figure.
\label{fig:Gamma-dep}}
\end{figure}

The comoving photon density $n_\gamma$ is decisive for both
1) the $\gamma\gamma$ optical depth $\tau_{\gamma\gamma}$
and hence the $\gamma\gamma$ cutoff energy $\varepsilon_{\gamma\gamma}$, and
2) the efficiency of $p\gamma$ interactions and hence the secondary cascade emission.
Figure \ref{fig:Esh-dep} displays single pulse spectra
for $\Gamma=300$, $\epsilon_{\rm B}/\epsilon_{\rm e}=1$, $\epsilon_{\rm p}/\epsilon_{\rm e}=10$, 
and varying pulse energies $E_{\rm sh}=10^{49}-10^{51}$ erg.
Higher $E_{\rm sh}$ implies higher $n_\gamma$,
and consequently stronger $p\gamma$ components as well as lower $\varepsilon_{\gamma\gamma}$.
Since $n_\gamma \propto \Gamma^{-5}$ with other parameters fixed, varying $\Gamma$ has larger effects.
Shown in Figure \ref{fig:Gamma-dep} are single pulse spectra 
for $E_{\rm sh}=10^{50}$ erg,  $\epsilon_{\rm B}/\epsilon_{\rm e}=1$, $\epsilon_{\rm p}/\epsilon_{\rm e}=30$
and $\Gamma=100-1000$.
$\Gamma=100$ allows a high $\varepsilon_{\rm pk}$, cascade-dominated spectrum,
even though $\epsilon_{\rm p}/\epsilon_{\rm e}$ is 3 times less than the analogous case in Figure \ref{fig:fp-dep}.
Increasing $\Gamma$ leads to higher maximum energies and less cascade contribution.
The spectral hardening $\ga 0.1$ GeV for $\Gamma=300$
and $\ga 10$ GeV for $\Gamma=1000$ is due to secondary IC.

Thus high proton-dominance does not always result in conspicuous proton-induced emission
if $\Gamma$ is sufficiently high.
Conversely, the absence of hard, high-energy components
does not necessary rule out proton-dominated GRBs.
In fact, the conditions most favorable for contributing to UHECRs
is that they escape the source with minimal $p\gamma$ losses,
which corresponds roughly to the criterion
$\Gamma \ga 300 (\Delta t/0.1{\rm s})^{-0.3} (E_{\rm sh}/10^{51} {\rm erg})^{0.2}$ in our model (AI07).
On the other hand, $\Gamma$ can be observationally constrained
through its strong influence on $\varepsilon_{\gamma\gamma}$ \citep[e.g.][AI07]{lit01}.
Since the pulse energy $E_{\rm sh}$ and timescale $\Delta t$ are also measurable,
we may hope to identify bursts where $p\gamma$ losses are likely to be efficient,
and then constrain $\epsilon_{\rm p}/\epsilon_{\rm e}$ from the high-energy spectra,
although some degeneracy with $\epsilon_{\rm B}/\epsilon_{\rm e}$ will remain.

\begin{figure}[htb!]
\centering
\epsscale{1.0}
\plotone{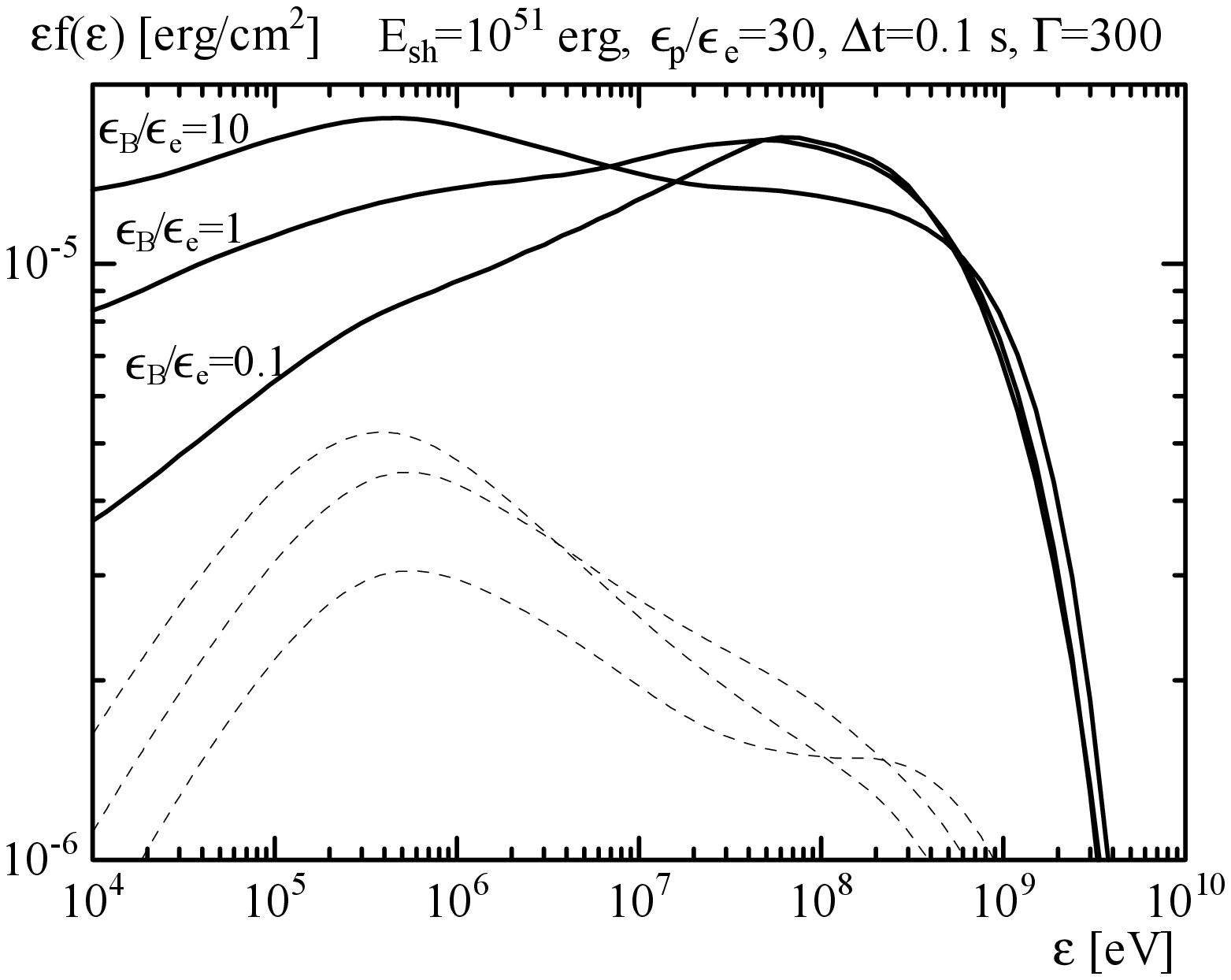}
\caption{
Single pulse, prompt photon spectra for
varying $\epsilon_{\rm B}/\epsilon_{\rm e}$ as labeled.
Other parameters are marked above the figure.
Dashed curves denote the primary components only,
whose peak flux decreases with $\epsilon_{\rm B}$.
\label{fig:fB-dep}}
\end{figure}

\begin{figure}[htb!]
\centering
\epsscale{1.0}
\plotone{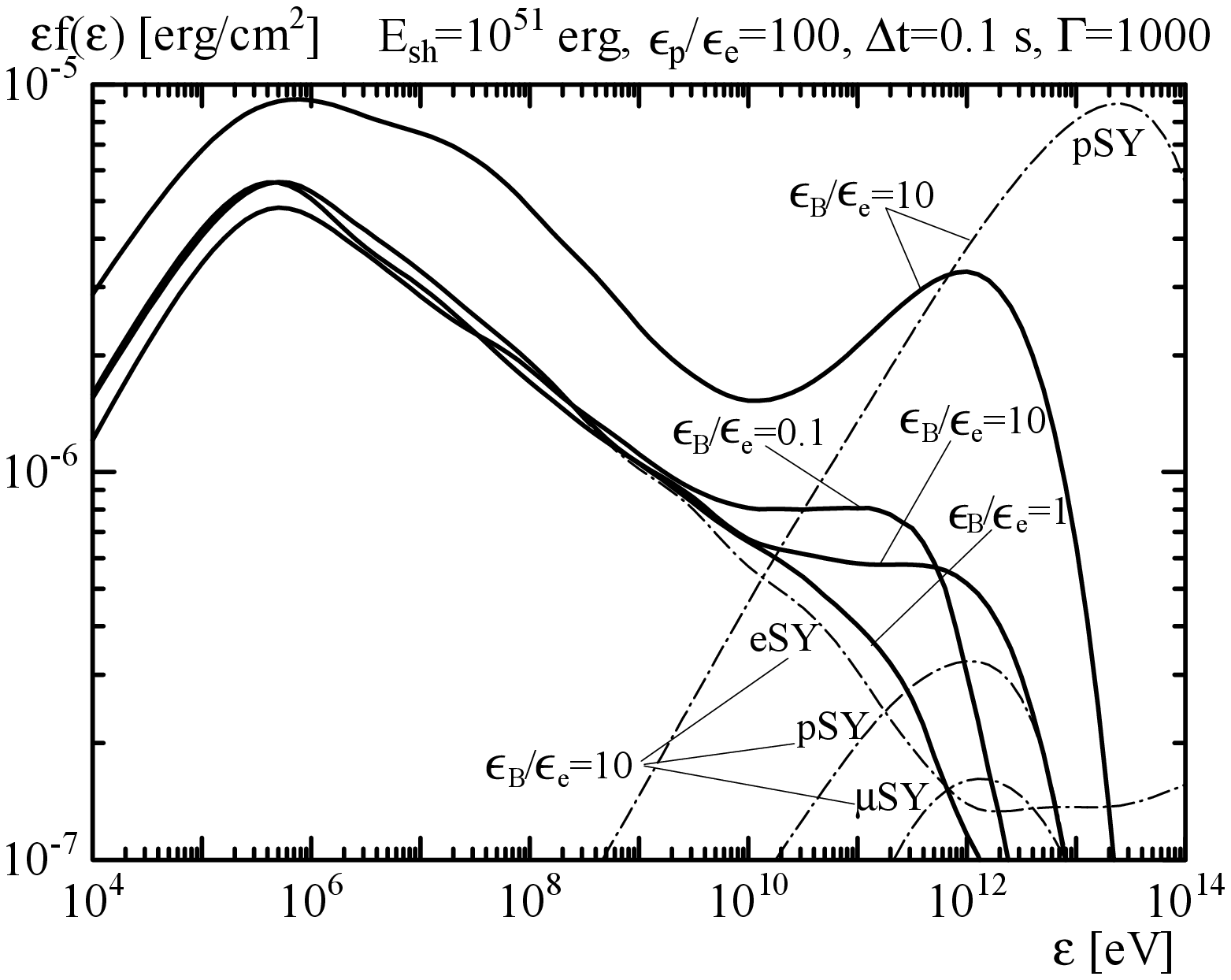}
\caption{
Single pulse, prompt photon spectra
varying $\epsilon_{\rm B}/\epsilon_{\rm e}$ as labeled.
Other parameters are marked above the figure.
Dot-dashed curves denote separately the electron synchrotron (eSY),
proton synchrotron (pSY) and muon synchrotron ($\mu$SY) components
without $\gamma\gamma$-absorption effects
for $\epsilon_{\rm B}/\epsilon_{\rm e}=10$ and 100.
\label{fig:fB-dep2}}
\end{figure}

Figure \ref{fig:fB-dep} shows single pulse spectra for
$E_{\rm sh}=10^{51}$ erg, $\Gamma=300$, $\epsilon_{\rm p}/\epsilon_{\rm e}=30$
and varying $\epsilon_{\rm B}/\epsilon_{\rm e}=0.1-10$.
The $\epsilon_{\rm B}/\epsilon_{\rm e}=1$ case is the same as in Figure \ref{fig:fp-dep}.
Higher $B$ causes steeper spectra with stronger secondary synchrotron relative to secondary IC,
while lower $B$ is vice-versa and produces a 100 MeV peak spectrum.
However, the dependence on $B$ can also be nontrivial.
In Figure \ref{fig:fB-dep2},
we show spectra for $E_{\rm sh}=10^{51}$ erg, $\Gamma=1000$,
$\epsilon_{\rm p}/\epsilon_{\rm e}=100$,
and varying $\epsilon_{\rm B}/\epsilon_{\rm e}=0.1-100$
(note that $\epsilon_{\rm B}/\epsilon_{\rm p} \le 1$).
The higher $\Gamma$ allows spectra extending into the TeV regime,
but renders $p\gamma$ processes inefficient despite the high proton-dominance.
All cases exhibit spectral bumps around 0.1-1 TeV, but their origins are quite different.
For $\epsilon_{\rm B}/\epsilon_{\rm e} \la 1$, this is due to secondary $e^\pm$ IC,
which is weaker for higher $B$.
However, when $\epsilon_{\rm B}/\epsilon_{\rm e} \ga 10$, the bump is stronger again,
owing to the appearance of synchrotron emission from protons and muons,
their ratio being roughly 2 to 1 for $\epsilon_{\rm B}/\epsilon_{\rm e}=10$
(dot-dashed curves in Figure \ref{fig:fB-dep2}).
For $\epsilon_{\rm B}/\epsilon_{\rm e}=100$,
we obtain a pronounced proton synchrotron TeV peak,
as well as enhanced emission at lower energies 
from synchrotron radiation by $e^\pm$ produced via $\gamma\gamma$ absorption.

\subsection{Equal Proton and Electron Indices}
\label{sec:equal}

\begin{figure}[htb!]
\centering
\epsscale{1.0}
\plotone{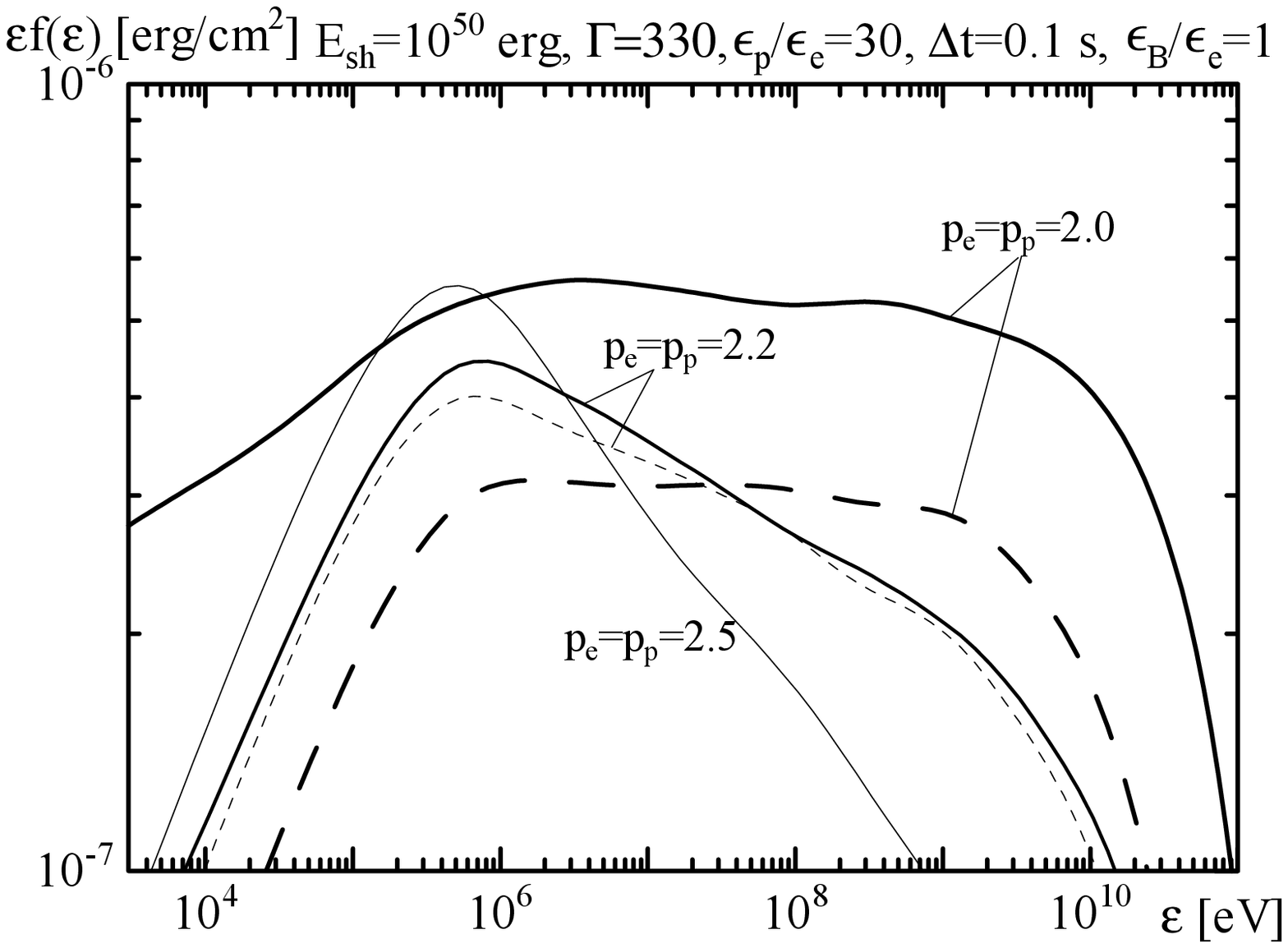}
\caption{
Single pulse, prompt photon spectra for varying values of $p_{\rm e}=p_{\rm p}$ as labeled.
Other parameters are marked above the figure.
Thick and thin dashed curves denote the primary components only,
for $p_{\rm e}=p_{\rm p}$=2.0 and 2.2, respectively.
\label{fig:index}}
\end{figure}

We now consider situations with $p_{\rm e}=p_{\rm p}$,
as would occur if the proton spectrum was a single power-law over its entire energy range.
Similar to the above,
Figure \ref{fig:index} testifies that
the spectrum for $\epsilon_{\rm p}/\epsilon_{\rm e}$=30 and $p_{\rm e}=p_{\rm p}$=2.0
can result in a hard GRB with photon index $\sim 2$ up to 10 GeV.
It is interesting to note that in such cases,
the spectral shape around the MeV peak alone
may not always reveal the correct value of $p_{\rm e}$.
However, for $p_{\rm e}=p_{\rm p}=2.2$,
the fraction of UHE protons and the associated cascade emission
is greatly diminished, except for a slight distortion of the spectrum above 100 MeV.
The proton contribution becomes totally negligible for $p_{\rm e}=p_{\rm p}=2.5$,
for which neither UHECRs nor neutrinos are significantly generated at any rate.

\section{Observational Implications}
\label{sec:obs}

A unique property of proton-dominated GRBs is that their photon spectra
can sometimes manifest very high peak energies in the 10 MeV-1 GeV range
due to $p\gamma$ cascade emission
(Figures \ref{fig:fp-dep},\ref{fig:Gamma-dep},\ref{fig:fB-dep},\ref{fig:index}).
This seems at variance with commonly observed values of
$\varepsilon_{\rm pk} \sim$ 0.1-1 MeV \citep{kan06}.
However, through a recent re-analysis of EGRET TASC data,
\citet{kan08} reported a GRB with apparently very high $\varepsilon_{\rm pk} > 170$ MeV,
as well as a few others with significant high-energy excess \citep[see also][]{gon03}.
Some studies have also indicated potential observational biases against BATSE detections
of high $\varepsilon_{\rm pk}$ \citep{llo99}.
At this moment, it is unclear how often such high $\varepsilon_{\rm pk}$ bursts occur,
and whether they are relevant to the proton-dominated cases discussed here,
or simply reflect a primary synchrotron peak energy that is much higher than average
(rather than the values we have assumed here).
In any case, the existence and nature of such bursts
will be definitively probed through ongoing observations
by {\it Fermi} \citep{omo06} and AGILE \citep{lon07}.
Note that it is also conceivable that some GRBs possess
conservative proton energies, say $E_{\rm p} \sim 10^{53}$ erg,
but with $E_{\rm p}/E_{\rm e} \gg 1$ so that the MeV emission is relatively weak. 
Even if unimportant for UHECRs (\S \ref{sec:intro}),
new generation satellites should also probe such MeV-weak bursts.

The $p\gamma$ cascade can also induce excess low-energy emission
(Figures \ref{fig:fp-dep},\ref{fig:Esh-dep},\ref{fig:fB-dep}),
which do not seem typical of known GRBs.
However, they may be relevant for some BATSE bursts
with soft excess components \citep{pre96},
or possibly a fraction of the X-ray rich GRBs \citep{sak05}.
{\it Fermi} and AGILE observations of the accompanying high-energy
excess will provide a test.

TeV detections of GRBs have yet to be achieved \cite[e.g.][]{atk05,alb07,hor07,aha09},
but some of the components discussed above may be eventually observed
by current ground-based facilities such as
MAGIC (II), HESS (II), VERITAS, CANGAROO III,
or the future projects CTA, AGIS, HAWC, etc.
For example, MAGIC may detect the luminous proton synchrotron emission
for $\epsilon_{\rm B}/\epsilon_{\rm e}=100$ in Figure \ref{fig:fB-dep2}
at 0.1 TeV beyond $z \sim 1$, assuming $E_{\gamma,{\rm iso}}=10^{53}$ erg
and the latest estimates of intergalactic $\gamma\gamma$ absorption \citep{alb08}.

Distinguishing between primary electron IC
and proton-induced emission components may not be easy from the spectral shape alone.
However, since the synchrotron and/or photomeson cooling timescales for UHE protons
are considerably longer than the cooling timescales for GeV-TeV emitting primary electrons,
we can expect important differences in their variability properties,
which should provide further observational clues.
Although this work was limited to time-averaged pulse spectra,
a desirable next step is to perform explicitly time-dependent calculations.

\section{Conclusions and Outlook}
\label{sec:conc}

Proton-dominated GRBs are motivated by physical considerations of particle acceleration
in collisionless shocks, as well as their potential to be the origin of UHECRs. 
In GRB UHECR scenarios, the spectral index for protons at UHE
must generally be harder than the typical indices for electrons emitting
in the multi-MeV range,
which may be possible depending on the physics of particle acceleration, cooling
and/or shock formation, as discussed in \S \ref{sec:model}.
Characteristic emission signatures can then result, such as
high peak energy bursts and/or excess low-energy emission
from photomeson-triggered pair cascades,
or luminous spectral bumps from proton synchrotron emission.
If the indices for electrons and protons at the respective energies are equal,
proton-related components may still be visible as long as the index $\la 2.2$,
but not for steeper spectra.
Through detailed observations of spectra and variability,
we may hope to disentangle the proton-induced components
from the competing emission process of inverse Compton from primary electrons.

Other observable consequences of proton-dominated GRBs
may include contributions to Galactic CRs \cite[e.g.][]{wic04}
and the diffuse high-energy neutrino background \citep[e.g.][]{mur07}.

We note that if some GRBs actually emit stronger GeV-TeV components
than previously expected as discussed here,
they could play an increased role
in probing high-$z$ intergalactic radiation fields (Inoue et al., in prep.)
as well as intergalactic magnetic fields  \cite[][and references therein]{ich08}.

\begin{acknowledgments}
We thank Chuck Dermer for very informative correspondence,
and Kohta Murase for valuable comments.
Support is acknowledged from
NSF PHY 0757155, NASA NNX08AL40G,
Grants-in-Aid for Scientific Research Nos. 19047004 and 19540283,
as well as the Global COE Program 
"The Next Generation of Physics, Spun from Universality and Emergence"
from the Ministry of E.C.S.S.T. (MEXT) of Japan.
\end{acknowledgments}


\end{document}